\documentclass[fleqn,twoside]{article}
\usepackage[headings]{espcrc2}
\usepackage{graphicx}
\usepackage[figuresright]{rotating}
\usepackage{epsfig}

\newcommand{\AmS}{{\protect\the\textfont2
  A\kern-.1667em\lower.5ex\hbox{M}\kern-.125emS}}

\title{Single top quark photoproduction at the LHC}

\author{J. de Favereau de Jeneret\address[UCL-CP3]{Universit\'e Catholique de Louvain, Center for Particle Physics and Phenomenology, Louvain-la-Neuve, Belgium}, S. Ovyn\addressmark[UCL-CP3] }

\begin{document}

\begin{abstract}
High-energy photon-proton interactions at the LHC offer interesting possibilities for the study of the electroweak sector up to TeV scale and searches for processes beyond the Standard Model. An analysis of the W associated single top photoproduction has been performed using the adapted MadGraph/MadEvent\cite{mgme} and CalcHEP\cite{calchep} programs interfaced to the Pythia\cite{pythia} generator and a fast detector simulation program. Event selection and suppression of main backgrounds have been studied. A comparable sensitivity to $|V_{tb}|$ to those obtained using the standard single top production in pp collisions has been achieved already for 10 fb$^{-1}$ of integrated luminosity. Photoproduction at the LHC provides also an attractive framework for observation of the anomalous production of single top due to Flavour-Changing Neutral Currents. The sensitivity to anomalous coupling parameters, $k_{tu\gamma}$ and $k_{tc\gamma}$ is presented and indicates that stronger limits can be placed on anomalous couplings after 1 fb$^{-1}$.
\end{abstract}

\maketitle

\section{Motivation}

The top quark is, because of its high mass, one of the least well known particles of the standard model. Among the properties that still have to be measured lie the electric charge and the $|V_{tb}|$ CKM matrix element. To measure the first, the production of the top quark should happen trough photon interaction in order to probe the photon-top coupling. The second requires a weak interaction trough the W boson which results in the production of a single top quark. 

Photoproduction at the LHC offers good conditions for both studies. First, the top quark photoproduction cross-section is important (around 2.5 pb). Moreover, more than 50$\%$ of this cross-section implies only one top quark in the final state, against only 5$\%$ in the case of partonic interactions. Second, an important part of this top production happens trough a photon-quark coupling, implying an important dependance of the cross-section in the top charge.

Probing the possible anomalous photoproduction of single top via flavour-changing neutral currents (FCNC) is also favoured at the LHC because of the high expected cross-section compared to the one expected at HERA. For same values of the anomalous coupling $k_{tu\gamma}$, the cross-section is expected to be two orders of magnitude higher. Furthermore, while at HERA only the up quark content of the proton contributed, the energy of the LHC allows to probe the proton at lower momentum fraction, opening the opportunity to probe the effect of the c quark via the $k_{tc\gamma}$ coupling.

\section{Standard Model single top photoproduction}

\begin{figure}
\epsfig{file=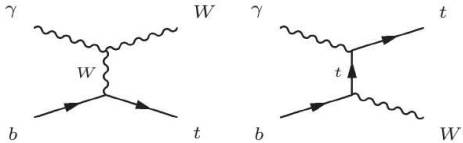,width=0.85\linewidth}
\caption{Diagrams for the dominant contribution to the SM production of single top quark.}
\label{sm_stop_diags}
\end{figure}

In the Standard Model (SM), single top photoproduction occurs at tree-level trough two main processes for which the diagrams can be seen at fig.~\ref{sm_stop_diags}. As two W bosons appear in the final state after the decay of the top quark, three type of topologies are possible, either fully leptonic, semileptonic or hadronic. Only the fully leptonic and semileptonic ones have been investigated, for which the cross-sections are 104 fb and 440 fb respectively. 

\subsection{Backgrounds}

All processes simulations have been done in the way described in \cite{sevtalk}. This includes tree-level computation of processes, hadronisation and a fast detector simulation based on particle 4-vector smearing and a cone algorithm for jets reconstruction.

Most of the backgrounds appear in two kind of processes. The first, called \textit{irreducible} comes from photoproduction with very similar final state as the signal. The second has the same final state but occurs through different processes induced by partonic interactions and is called \textit{reducible background}.

Main backgrounds for the leptonic topology come from processes including two W bosons and either a b-jet or a light jet that could be mistagged as a b-jet. These include $t\bar{t}$ for which one jet is not reconstructed. The semileptonic topology has more contributing backgrounds due to the various possible jet production processes. These mainly include semileptonic and leptonic $t\bar{t}$ production and W boson production associated with three jets or more. This is summarized in tables \ref{lep_bkg} and \ref{sem_bkg}. 

One should note that inelastic photoproduction, namely the case for which the proton having emitted a photon does not survive the interaction, has not been taken into account. This would add to the cross-section of both signal and irreducible backgrounds. The cross-sections for such inelastic processes is not precisely known, as because of the small impact parameter the probability of rescattering becomes important. This makes the efficiency of tagging such events much harder to compute precisely, leading to important systematic errors. Diffractive background processes have also not been considered here, although they can look very similar to photoproduction in some cases.

\begin{table}
\renewcommand{\arraystretch}{1.2}
\caption{Background processes used in the leptonic channel. Cross-sections include the branching ratio of the W boson to any lepton and generation cuts of $P_T > 10~\textrm{GeV}$ for leptons and $P_T > 20~\textrm{GeV}$ for jets.}
\begin{tabular}{l@{$~\rightarrow~$}lr@{}lc}
\hline
\multicolumn{2}{c}{process} & \multicolumn{2}{c}{$\sigma$ [fb]} & sample size \\
\hline
$\gamma$p & $t\bar{t}$(leptonic)  & 159 &                & 100 k       \\
$\gamma$p & $W^+ W^- q'$          & 63  &                &  90 k       \\
pp        & $t\bar{t}$            & 73  & $\times 10^3$  & 510 k       \\
pp        & $W^+ W^- j$           & 5.2 & $\times 10^3$  &  50 k       \\
\hline
\end{tabular}
\label{lep_bkg}
\end{table}

\begin{table}
\renewcommand{\arraystretch}{1.2}
\caption{Background processes used in the semileptonic channel. Cross-sections include the branching ratio of the W boson to any lepton and generation cuts of $P_T > 10~\textrm{GeV}$ for leptons and $P_T > 20~\textrm{GeV}$ for jets.}
\begin{tabular}{l@{$~\rightarrow~$}lr@{}lc}
\hline
\multicolumn{2}{c}{process} & \multicolumn{2}{c}{$\sigma$ [fb]} & sample size \\
\hline
$\gamma$p & $t\bar{t}$ (1l,2l) & 831 &                & 270 k       \\
$\gamma$p & $W^+ W^- jjj$      & 2.8 & $\times 10^3$  &  50 k       \\
$\gamma$p & $W b\bar{b} j$     & 55  &                &  50 k       \\
pp        & $t\bar{t}$ (1l,2l) & 407 & $\times 10^3$  & 520 k       \\
pp        & $W n\times j$      &  73 & $\times 10^6$  & 770 k       \\
pp        & $W b\bar{b} j$     & 267 & $\times 10^3$  & 120 k       \\
pp        & $tj$               &  67 & $\times 10^3$  & 100 k       \\ 
\hline
\end{tabular}
\label{sem_bkg}
\end{table}

\subsection{Partonic backgrounds rejection}
\label{ppreject}

The key difference between photoproduction and partonic interactions at the LHC lies in the absence of colour exchange on the photon side. This causes an important zone of rapidity to be completely devoid of hadronic activity, while the colour flow in partonic interactions tends to produce hadrons in the region between the proton remnant and the hard hadronic final states. This region with no hadronic activity is usually called a \textit{large rapidity gap} (LRG) and is a natural way to distinguish between photoproduction and partonic backgrounds. 

In the case of this study, the LRG requirement takes the following shape : the side of the detector with the minimum forward activity is determined from forward calorimeters taking place between $|\eta|$ = 3 and $|\eta|$ = 5 and this side is tagged as the \textit{gap side} or \textit{photon side}. Then, a cut is placed on the energy in the forward calorimeter of the gap side, discriminating between partonic processes for which this energy is high and photoproduction where it is usualy very close to zero. This method can obviously be used only when the instantaneous luminosity is low, as higher luminosities imply the appearance of pileup events that immediately fill the gap.

\begin{figure}
\epsfig{file=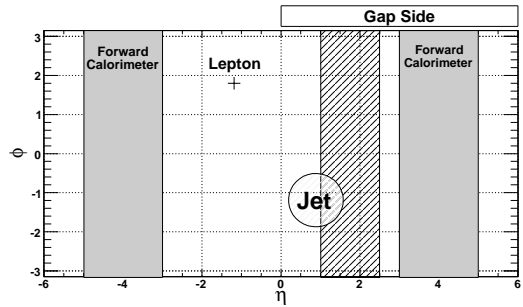,width=1.05\linewidth}
\caption{Illustration of the LRG and exclusivity conditions used in order to reduce the impact of partonic backgrounds. Grey areas represent the forward calorimeters, among which the one having measured the lowest energy defines the gap side. The shaded area, choosen to be on the gap side, represents the area in which tracks are counted for the exclusivity condition. One jet cone has been removed from this zone as tracks from the jet should not enter into the exclusivity requirement.}
\label{gap_exclu}
\end{figure}

Another tagging method based on the same physics properties of photoproduction events is to place an \textit{exclusivity condition} on reconstructed particle tracks on the gap side. This means that a fixed central region of the tracking device should be empty of tracks, excluding lepton tracks and jet cones. This is illustrated by fig.~\ref{gap_exclu}. This condition has the importat advantage to be useable in cases of higher luminosities as well if proper vertex determination is possible, as one can take into account only tracks issued from the same vertex as the studied final state. Those methods are described in details in \cite{sevtalk}.

Unfortunately, when the rapidity gap condition cannot be used, the exclusivity condition alone cannot reduce patronic backgrounds to a level that allows proper signal extraction. In that case, elastic photon emission can be tagged only using \textit{very forward detectors} (VFD) placed hundreds of meters away from the interaction point. Those detect protons having loss a sufficient amount of energy as they are deflected in a different way by beam magnets that act as spectrometers in that case. For a complete description of this method, see \cite{xavtalk}. 

\begin{figure}
\epsfig{file=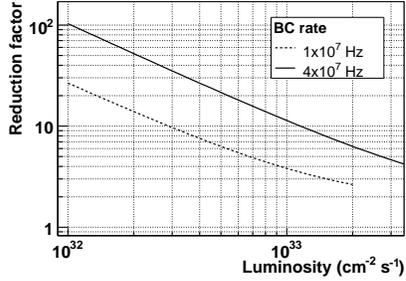,width=0.8\linewidth}
\label{rej_power}
\caption{Rejection power of partonic backgrounds as a function of the instantaneous luminosity for two different bunch-crossing rates scenarii. This factor can directly be applied to the cross-section in order to get its contribution after the requirement of at least one VFD hit.}
\end{figure}

However, using VFDs cannot provide a total rejection of the partonic processes because of the presence of single diffractive events in the pileup. Those events have a very high cross-section ($\sim 10 \textrm{mb}^{-1}$) and also contain a surviving proton in the final state. When such events take place in the pile-up, the overall event mimics well a photoproduction event. The probability of such accidental co\"incidences provides directly the rejection power of VFDs. For instance, the case for which VFD stations would be put at 220m and 420m from the interaction point has been computed and provides rejection factors of 11 at $10^{33} \textrm{cm}^{-2} \textrm{s}^{-1}$ and 5.6 at $2 \times 10^{33} \textrm{cm}^{-2} \textrm{s}^{-1}$, as can be seen on fig.~\ref{rej_power}.

\subsection{Signal selection}

The overall selection of any photoproduction signal consists of three steps : first, topology-based cuts are applied, then one should ensure that partonic backgrounds are reduced to the same level as photoproduction ones and eventually more advanced cuts based on kinematics are applied to reach an optimal signal over background ratio.

To select the semileptonic signal, the following cuts have been applied : 

\begin{itemize}
 \item three jets with $P_T > 30~\textrm{GeV}$,
 \item one lepton with $P_T > 20~\textrm{GeV}$,
 \item energy in one forward calorimeter $<~30~\textrm{GeV}$,
 \item exclusivity condition on the region $1~<~\eta~<~2.5$, 
 \item one jet tagged as a b-jet,
 \item scalar sum of all $P_T~<~230~\textrm{GeV}$,
 \item reconstructed mass of non-b jets around W mass ($\Delta_M~< 20~\textrm{GeV}$).
\end{itemize}

After this selection, the cross-section for the signal is reduced to 4.8 fb, against 5.5 fb for the backgrounds, 65 $\%$ of which comes from partonic processes. Details are given in table~\ref{sel_semi}.

\begin{table}
\caption{Effect of various cuts on the cross-section of the semileptonic signal, photoproduction backgrounds and partonic backgrounds.}
\renewcommand{\arraystretch}{1.2}
\begin{tabular}{lccc}
\hline
$\sigma$ [fb] & signal & $\gamma$p & pp \\
\hline
production    & 440.0 & 3.6 $\times 10^3$ & 74 $\times 10^6$  \\
topology cuts &  36.0 & 144.4             & 116 $\times 10^3$ \\
gap + exclu.  &  24.2 & 77.9              & 187.5             \\
\hline
               &     & 1.9               & 3.6               \\
\raisebox{1.5ex}{final cuts} & \raisebox{1.5ex}{4.8} &\multicolumn{2}{c}{5.5}              \\
\hline
\end{tabular}
\label{sel_semi}
\end{table}

The selection for the leptonic channel consists of the following cuts : 

\begin{itemize}
 \item one jet with $P_T~>~30~\textrm{GeV}$,
 \item two leptons with $P_T~>~20~\textrm{GeV}$,
 \item missing transverse energy $>~20~\textrm{GeV}$,
 \item Energy in one forward calorimeter $<~30~\textrm{GeV}$,
 \item Exclusivity condition on the region $1~<~\eta~<~2.5$, 
 \item the jet tagged as a b-jet.
\end{itemize}

Signal cross-section for this topology is 4.9 fb after cuts, for a background cross-section of 2.2 fb with less than 30 $\%$ of partonic contribution. Details are in table~\ref{sel_lep}.

\begin{table}
\caption{Effect of various cuts on the cross-section of the leptonic signal, photoproduction backgrounds and partonic backgrounds.}
\renewcommand{\arraystretch}{1.2}
 \begin{tabular}{lccc}
\hline
$\sigma$ [fb] & signal & $\gamma$p & pp \\
\hline
\rule{0pt}{3ex} production    & 104.0 & 222    & 83 $\times 10^3$  \\
topology cuts &  14.2 & 13.7   & 3.4 $\times 10^3$ \\
gap + exclu.  &  12.7 &  8.0   & 3.2               \\
\hline
              &     &   1.6    &  0.6              \\
\raisebox{1.5ex}{final cuts}  & \raisebox{1.5ex}{4.9} &\multicolumn{2}{c}{2.2}      \\
\hline
\end{tabular}
\label{sel_lep}
\end{table}

\subsection{Systematic errors}

Various sources of systematic errors have been investigated : 

\begin{itemize}
 \item{Jet energy scale (JES) :} Jet(s) energy have been scaled up and down by 5 $\%$ for jets with $P_T < 30~\textrm{GeV}$, 3 $\%$ for jets with $P_T > 50~\textrm{GeV}$ and a linear interpolation between these two boundaries.
 \item{Exclusivity :} The track reconstruction efficiency, fixed to 90 $\%$ by default, has been moved to 85 $\%$ and 95 $\%$.
 \item{Rapidity gap :} The cut on the energy in the forward calorimeter of the gap side has been moved by 10 $\%$ upwards and downwards.
 \item{Luminosity :} An overall luminosity uncertainty of 5 $\%$ has been assumed.
 \item{b-tagging :} An uncertainty of 5 $\%$ has been assumed on the b-tagging of a b-jet. No error on mis-tagging was assumed.
 \item{Theoretical uncertainty :} The error is process-dependant. When no estimate was found in the litterature for photoproduction processes, the same uncertainty as for the corresponding partonic process was taken for a pessimistic estimate. Partonic cross-sections after cuts have been considered known to the 2 $\%$ level as the cross-section without application of the rapidity gap and exclusivity conditions can be measured directly and the error on the effect of these cuts is computed separately.
\end{itemize}

The effect of these can be seen on tables \ref{sys_lep} and \ref{sys_sem}. The total error is dominated by the rapidity gap and exclusivity cut errors.

\begin{table}
 \caption{Systematic errors on signal and backgrounds for the leptonic topology.}
 \renewcommand{\arraystretch}{1.2}
 \begin{tabular}{lcc}
  \hline
Error            & signal ($\%$) & background ($\%$) \\
Jet energy scale & 0.6           & 3.7 \\
Rapidity gap     & 0.8           & 3.0 \\
Exclusivity      & 1.4           & 7.9 \\
Luminosity       & 5.0           & 5.0 \\
Theoretical      & 6.0           & 3.4 \\
b-tagging        & 5.0           & 0.0 \\
\hline
total            & 9.4           & 11.0 \\
\hline 
 \end{tabular}
\label{sys_lep}
\end{table}

\begin{table}
 \caption{Systematic errors on signal and backgrounds for the semileptonic topology.}
 \renewcommand{\arraystretch}{1.2}
 \begin{tabular}{lcc}
  \hline
Error            & signal ($\%$) & background ($\%$) \\
Jet energy scale & 6.7           & 10.6 \\
Rapidity gap     & 0.5           & 12.5 \\
Exclusivity      & 1.2           & 2.6 \\
Luminosity       & 5.0           & 5.0 \\
Theoretical      & 6.0           & 2.0 \\
b-tagging        & 5.0           & 0.0 \\
\hline
total            & 11.5          & 17.5 \\
\hline 
 \end{tabular}
\label{sys_sem}
\end{table}

\subsection{Results}

The total error on the measured cross-section is given by the following formula :

\begin{eqnarray}
\frac{\Delta \sigma_{obs}}{\sigma_{obs}} = \frac{\Delta \varepsilon}{\varepsilon} \oplus \frac{\Delta L}{L} \oplus \left[ \frac{B}{S} \right] \frac{\Delta B}{B}  \nonumber \\ \oplus \left[ \frac{B}{S} + 1 \right] \frac{\Delta N}{N} 
\end{eqnarray}

Where $\Delta \varepsilon$, $\Delta L$ and $\Delta B$ are the systematic errors estimates on the signal selection efficiency, the luminosity and the background cross-section respectively and $\Delta N$ is the statistical error on the observed number of events. The different contributions are given in table~\ref{sys_final_sm}.

\begin{table}
\caption{Contributions to the total cross-section measurement error.}
\renewcommand{\arraystretch}{1.5}
\begin{tabular}{lcc}
\hline
Error                                                   & leptonic [$\%$]&  semileptonic [$\%$]\\
\hline
$\frac{\Delta \varepsilon}{\varepsilon}$                &  5.3 &  8.5 \\
$\frac{\Delta L}{L}$                                    &  5.0 &  5.0 \\
$\left[ \frac{B}{S} \right] \frac{\Delta B}{B}$         &  6.4 & 17.3 \\
$\left[ \frac{B}{S} + 1 \right] \frac{\Delta N}{N}$     & 17.4 & 17.9 \\
\hline
total                                                   & 19.4 & 33.3 \\
\hline
\end{tabular}
\label{sys_final_sm}
\end{table}

From the error on the total single top cross-section measurement, one can compute the error on the $|V_{tb}|$ measurement from the formula : 

\begin{equation}
\frac{\Delta |V_{tb}|}{|V_{tb}|} = \frac{1}{2} \left[ \frac{\Delta \sigma_{obs.}}{\sigma} \oplus \frac{\Delta \sigma_{theo.}}{\sigma} \right]
\end{equation}

The expected error on the measurement of $|V_{tb}|$ is 16.9 $\%$ for the semileptonic channel and 10.1 $\%$ for the leptonic one after 10 fb$^{-1}$ of integrated luminosity, while the expected uncertainty from the equivalent study based on partonic interactions is 14 $\%$ \cite{vtb_pp} using the same integrated luminosity, showing that photoproduction is at least competitive with partonic-based studies and that the combination of both studies could lead to significant improvement of the error.

\section{Anomalous single top photoproduction}

\begin{figure}
\epsfig{file=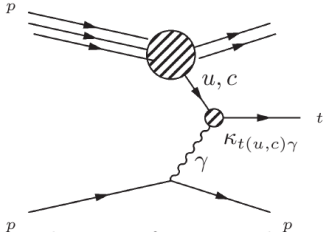,width=0.6\linewidth}
\label{anotop_diagram}
\caption{Main diagram for FCNC production of single top.}
\end{figure}

In the Standard model, single top photoproduction with no associated final states does not occur at tree-level. The observation of such a final state would be a direct hint of the existence of FCNC. The diagram for single top photoproduction via FCNC is shown on fig.~\ref{anotop_diagram}.

An effective lagrangian describing such interactions can be written as \cite{anotop_lag} :

\begin{eqnarray}
\mathcal{L}~=~iee_t\bar{t}\frac{\sigma_{\mu\nu}q^{\nu}}{\Lambda}k_{tu\gamma}uA^{\mu} & \nonumber \\ 
+~iee_t\bar{t}\frac{\sigma_{\mu\nu}q^{\nu}}{\Lambda}k_{tc\gamma}cA^{\mu} & +~h.c.,
\end{eqnarray}

where $\sigma_{\mu\nu}$ is defined as $(\gamma^{\mu} \gamma^{\nu} - \gamma^{\nu} \gamma^{\mu})/2$, $q^{\nu}$ being the photon 4-vector and $\Lambda$ an arbitrary scale,
conventionally taken as the top mass. The coulings k$_{tu\gamma}$ and k$_{tc\gamma}$ are real and positive. By introducing this lagrangian in CalcHep, the following cross-section was obtained as a function of the couplings : 

\begin{equation}
\sigma_{pp \rightarrow t} = 368~\textrm{pb}\times k^2_{tu\gamma} + 122~\textrm{pb}\times k^2_{tc\gamma}.
\end{equation}

The best actual limit on $k_{tu\gamma}$, obtained by the H1 collaboration at the HERA collider, is around 0.14 \cite{h1_ktug}, while the $k_{tc\gamma}$ coupling has not been probed yet. 

The studied final state consists of a leptonic decay of the W boson coming from the top quark, giving a final topology consisting of a hard lepton and a jet from the b quark.

\subsection{Backgrounds}

The dominant background processes for this final state come from events with one W boson and one jet (mis-)tagged as a b-jet. The b-tagging method is such that jets coming from c quarks have a high probability to be tagged as b-jets (around 10 $\%$) making it an important background, at the same level as the light jet contribution, for which the mistagging probability is only around 1 $\%$. The contribution of genuine b-jets is negligible because of the low cross-section of the process, which is three orders of magnitude lower than the cross-section of the W + c topology. Backgrounds cross-sections and sample sizes are given in table~\ref{ano_bkg}. 

\begin{table}
\renewcommand{\arraystretch}{1.2}
\caption{Background processes used for the analysis of the anomalous top photoproduction. Cross-sections include the branching ratio of the W boson to electron or muon and generation cuts of $P_T~>~10~\textrm{GeV}$ for leptons and $P_T~>~20~\textrm{GeV}$ for jets.}
\begin{tabular}{l@{$~\rightarrow~$}lr@{}lc}
\hline
\multicolumn{2}{c}{process} & \multicolumn{2}{c}{$\sigma$ [fb]} & sample size \\
\hline
$\gamma$p & $W j$  & 41.6 & $\times 10^3$ & 100 k       \\
$\gamma$p & $W c$  & 11.5 & $\times 10^3$ & 100 k       \\
pp        & $W j$  & 77.3 & $\times 10^6$ & 100 k      \\
pp        & $W c$  &  8.8 & $\times 10^6$ & 100 k      \\
\hline
\end{tabular}
\label{ano_bkg}
\end{table}

\subsection{Signal selection}

The overall selection consists of : 

\begin{itemize}
 \item one jet with $P_T~>~45~\textrm{GeV}$,
 \item one lepton with $P_T~>~20~\textrm{GeV}$,
 \item the jet tagged as a b-jet,
 \item Exclusivity condition on $1~<~\eta~<~2.5$,
 \item the reconstructed invariant mass of the top quark between 140 GeV and 210 GeV. 
\end{itemize}

This study has also been performed considering the presence of pileup preventing the use of a LRG-based selection. When the luminosity gets above $10^{33} \textrm{cm}^{-2} \textrm{s}^{-1}$, the effect of the pileup becomes important and we assumed the presence of VFD as described in the previous section. A proper simulation of the proton propagation in the LHC beamline performed using HECTOR \cite{hector} shows that using detectors stations at 220m and 420m frome the interaction point, one selects events for which the proton has lost between 20 GeV and 800 GeV. As stated before, the reduction of the partonic background is not strong enough to reach the same level as the photoproduction backgrounds. 

Another advantage of the VFD is that, considering a well designed reconstruction algorithm, one can deduce the energy loss of the proton that hits the detector and use it in order to improve the selection of photoproduction processes. This can easily be used here to constrain the reconstruction.

\begin{figure}
\epsfig{file=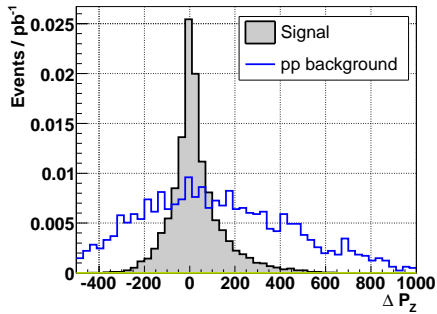,width=0.8\linewidth}
\caption{Distribution of the difference between the top quark longitudinal momentum reconstructed from the central detector and from a VFD. The distribution is shown for the anomalous top signal (full) and for the partonic background(empty).}
\label{del_pz}
\end{figure}

An additionnal cut was designed by reconstructing the top quark longitudinal momentum both from the central event and from the proton energy loss. The difference between these two values allows to distinguish between photoproduction events for which they are close, and partonic events for which the distance between them is distributed randomly. These distrubutions can be seen on fig.~\ref{del_pz}.

Eventually, the selection in the case of very low luminosities - for which the pileup is negligible - is the one described higher supplemented by a strong rapidity gap condition (energy in the forward calorimeter lower than 20 GeV), while at higher luminosity the VFDs have been used as well as the cut on the difference between the reconstructed longitudinal momentums described previously. 
%

\subsection{Systematic errors}

The same systematic uncertainties as in the case of the SM single top study have been estimated. One again, the rapidity gap and exclusivity condition account for the most important part of it. In the case of higher luminosities for which the VFDs where used, no systematic error was assumed on this tagging. The detail of all errors for both scenarii are given on table~\ref{sys_ano} . Differences between the two scenarii mainly come from the different amount of partonic background in the two selected samples. Signal systematics stay unaffected by the scenario change, as the error due to the LRG requirement is negligible.

\begin{table}
 \caption{Systematic errors on signal and backgrounds for both scenarii.}
 \renewcommand{\arraystretch}{1.2}
 \begin{tabular}{lccc}
  \hline
Error         & signal ($\%$) & \multicolumn{2}{c}{Background ($\%$)}\\
              &               & very low          & low              \\
\hline
JES           & 1.6           & 3.0               & 3.3 \\
LRG           & 0.0           & 9.9               &  -  \\
Exclusivity   & 1.0           & 5.5               & 6.9 \\
Luminosity    & 5.0           & 5.0               & 5.0 \\
Theoretical   & 5.0           & 1.9               & 1.3 \\
b-tagging     & 5.0           & 0.0               & 0.0 \\
\hline
total         & 8.9          & 12.9               & 9.3 \\
\hline 
 \end{tabular}
 \label{sys_ano}
\end{table}

\subsection{Results}

Using the LRG requirement for an integrated luminosity of 1 fb$^{-1}$, one gets the following final result ($k_{tu\gamma} = 0.15$, $k_{tc\gamma} = 0$) : 

\begin{eqnarray*}
\textrm{Signal : }     83.2 \pm 9.1 \textrm{(stat.)} \pm 7.4 \textrm{(syst.)~events} \\
\textrm{Background : } 12.7 \pm 3.6 \textrm{(stat.)} \pm 1.6 \textrm{(syst.)~events} \\
\end{eqnarray*}

While for 30 fb$^{-1}$ at higher luminosity using the VFDs to tag photoproduction the final sample is composed of : 

\begin{eqnarray*}
\textrm{Signal : }       1554 \pm 39 \textrm{(stat.)} \pm 138 \textrm{(syst.)~events} \\
\textrm{Background : }   327 \pm  18 \textrm{(stat.)} \pm 30  \textrm{(syst.)~events} \\
\end{eqnarray*}

In order to set a limit on the anomalous couplings, we assumed a measurement in agreement with the SM, i.e seeing the background only. Given this measurement, we computed the maximum cross-section for which this measurement was not less than 5$\%$ probable. This cross-section corresponds to the minimum real cross-section for which the ``SM only'' hypothesis will be rejected at 95 $\%$ C.L. and thus gives the minimum anomalous cross-section once the SM cross-section is substracted.

The number of events was assumed to be distributed according to a Poisson distribution, while the systematic uncertainty was included using Monte Carlo to obtain a realistic convolution of statistical and systematic errors.

The obtained expected limits on the anomalous couplings are the following : 

\begin{eqnarray*}
\textrm{Very low luminosity : } k_{tu\gamma} < 0.044,~ k_{tc\gamma} < 0.077, \\
\textrm{Low luminosity : } k_{tu\gamma} < 0.029,~ k_{tc\gamma} < 0.050. \\
\end{eqnarray*}

\section{Conclusions and prospects}

Both above study show the potential to bring precise measurement of important parameters linked to top quark physics. The $|V_{tb}|$ measurement with a precision of 10 $\%$ to 17 $\%$ should be competitive to the partonic processes-based one, allowing to improve it. The anomalous couplings limits should be at least 3-4 times better than the ones obtanied at HERA.

However, both studies will be refined when full detector simulation will be used, providing a better estimate of the systematic errors in order to replace the pessimistic ones used in the present analysis. Also, studying the influence of diffractive backgrounds is an important part of the work to be done, as well as the contribution of inelastic photon emissions to both signal and backgrounds.

\end{document}